\documentclass[12pt]{amsart}
\usepackage{amssymb}
\usepackage{hyperref}
\usepackage{amsmath}
\usepackage{amscd}
\usepackage{graphicx}

\DeclareMathSymbol{\shortminus}{\mathbin}{AMSa}{"39}

\newcommand{\group}[1]{\mathrm{#1}}

\newcommand{\rep}[1]{\rep{#1}}

\newcommand{\Aut}{\operatorname{Aut}}

\newcommand{\mon}{\vec{H}}

\newcommand{\Tr}{\operatorname{Tr}}

\newcommand{\End}{\operatorname{End}}

\newcommand{\C}{\mathbb{C}}

\newcommand{\N}{\mathbb{N}}

\newcommand{\V}{\mathbf{V}}
\newcommand{\W}{\mathbf{W}}
\newcommand{\Z}{\mathbb{Z}}

\newcommand{\Y}{\mathsf{Y}}

\newcommand{\BK}{\mathrm{BK}}

\newcommand{\IZ}{\mathrm{IZ}}

\newtheorem{thm}{Theorem}[section]
\newtheorem{prop}[thm]{Proposition}
\newtheorem{lem}[thm]{Lemma}

\theoremstyle{definition}

\title[Berezin-Karpelevich Integral]{Combinatorics of the Berezin-Karpelevich Integral}

\author{Jonathan Novak}

\begin{document}

\begin{abstract}
	The Berezin-Karpelevich integral is a double integral over unitary 
	matrices which plays the role of the Itzykson-Zuber integral in 
	rectangular matrix models.  We obtain a topological expansion of the 
	Berezin-Karpelevich integral in terms of monotone Hurwitz numbers,
	and obtain from this certain combinatorial identities.
\end{abstract}

\maketitle

\section{Introduction}

\subsection{Itzykson-Zuber integral}
The Itzykson-Zuber integral,

	\begin{equation}
	\label{eqn:IZ}
		I_N = \int_{\group{U}_N} e^{z\Tr AUBU^*} \mathrm{d}U,
	\end{equation}
	
\noindent
is a unitary matrix integral which first appeared in mathematical physics 
in the context of multimatrix models \cite{IZ}. It depends on a coupling
parameter $z \in \C$ and a pair of matrices $A,B \in \C^{N \times N}$, 
with the dependence on these matrices only through their eigenvalues 
$a_1,\dots,a_N,b_1,\dots,b_N \in \C$. Thus, $I_N$ is an entire function 
of $1+2N$ complex variables. Since $I_N$ is invariant under independent permutations of 
$a_1,\dots,a_N$ and $b_1,\dots,b_N$, and is stable under swapping these two sets of variables, 
its Maclaurin series can be presented in the form

	\begin{equation}
	\label{eqn:StringExpansionIZ}
		I_N = 1 + \sum_{d=1}^\infty \frac{z^d}{d!} \sum_{\alpha,\beta \in \Y^d} 
		\frac{p_\alpha(a_1,\dots,a_N)}{N^{\ell(\alpha)}} \frac{p_\beta(b_1,\dots,b_N)}{N^{\ell(\beta)}}
		I_N(\alpha,\beta),
	\end{equation}
	
\noindent
where the internal sum is over pairs of Young diagrams $\alpha,\beta$ each consisting of $d$ cells and
$p_\alpha,p_\beta$ are the corresponding Newton symmetric polynomials, each of which
we have normalized by its maximum modulus on the unit polydisc in $\C^N$. We call this the \emph{string expansion}
of the Itzykson-Zuber integral, and the numbers $I_N(\alpha,\beta)=I_N(\beta,\alpha)$ its
\emph{string coefficients}. We also have a \emph{connected string expansion}

	\begin{equation}
	\label{eqn:ConnectedStringExpansionIZ}
		\log I_N = \sum_{d=1}^\infty \frac{z^d}{d!} 
		\sum_{\alpha,\beta \in \Y^d} \frac{p_\alpha(a_1,\dots,a_N)}{N^{\ell(\alpha)}} \frac{p_\beta(b_1,\dots,b_N)}{N^{\ell(\beta)}}
		L_N(\alpha,\beta),
	\end{equation}
	
\noindent
in which $L_N(\alpha,\beta)=L_N(\beta,\alpha)$ are the \emph{connected string coefficients} 
of the Itzykson-Zuber integral and the series converges absolutely in a 
neighborhood of the origin in $\C^{1+2N}$.

The quest for a topological expansion of the string coefficients of $I_N$  
began in \cite{IZ}, where it was shown that the limits

	\begin{equation}
	\label{eqn:LimitsIZ}
		L(\alpha,\beta) = \lim_{N \to \infty} N^{d-2} L_N(\alpha,\beta)
	\end{equation}  
	
\noindent
exist and are integers for all Young diagrams $\alpha,\beta$ with $d \leq 8$. 
By analogy with the planar limit of the Hermitian one-matrix model \cite{BIPZ},
this integrality leads to the hypothesis that $L(\alpha,\beta)$ counts some class of ``genus zero'' combinatorial
objects. More ambitiously, one hopes for the existence of subleading corrections which enumerate objects of higher genus,
as in with Hermitian matrix integals \cite{BIZ}. However, due to the non-Gaussian nature of Haar measure,
it is a mistake to insist that the combinatorics of the Itzykson-Zuber integral must be
expressed in terms of maps on surfaces \cite{ZZ}.

The combinatorial structure underlying the Ityzkson-Zuber integral was exposed in \cite{GGN3}, where map counting 
was replaced by its older sibling, the enumeration of branched covers of Riemann surfaces \cite{Lando:ICM}.
Let us identify the symmetric group $\group{S}^d=\Aut\{1,\dots,d\}$ with its
Cayley graph as generated by the conjugacy class of transpositions. Write $W^r(\alpha,\beta)$
for the number of $r$-step walks on $\group{S}^d$ which begin at a permutation of cycle type 
$\alpha$ and end at a permutation of cycle type $\beta$. 
The counting function $W^r(\alpha,\beta)$ also enumerates
degree $d$ branched covers of the Riemann sphere with $r$ simple branch points and two
additional points over which the covering map has ramification profiles $\alpha$ and $\beta$. 
By the Riemann-Hurwitz formula,
$W^r(\alpha,\beta)=0$ unless $r=2g-2+\ell(\alpha)+\ell(\beta)$ where $g \in \Z$ is the 
genus of the possibly disconnected covering surface. We write $W^r(\alpha,\beta)=H_g^\bullet(\alpha,\beta)$
when the Riemann-Hurwitz constraint is satisfied and call $H_g^\bullet(\alpha,\beta)$ a 
\emph{disconnected double Hurwitz number} of genus $g$. The corresponding \emph{connected double Hurwitz number}
$H_g(\alpha,\beta)$ counts walks whose steps and endpoints generate a transitive subgroup of the symmetric group, 
or equivalently connected covers. The double Hurwitz numbers $H_g^\bullet(\alpha,\beta)$ and 
$H_g(\alpha,\beta)$ were first studied by Okounkov \cite{MRL}. Hurwitz
himself had considered the numbers $H_g^\bullet(\alpha)$ and $H_g(\alpha)$ enumerating branched covers
with just one point of prescribed ramification.
 
Now let us enrich the Cayley graph by marking
each edge of $\group{S}^d$ corresponding to the transposition $(i\ j)$
with the larger symbol $j$. This is the \emph{Jucys-Murphy labeling}
of the symmetric group \cite{Jucys,Murphy}, which proves to be a useful device \cite{DG,OV}. The \emph{monotone double Hurwitz numbers} 
$\mon_g^\bullet(\alpha,\beta)$ and $\mon_g(\alpha,\beta)$, introduced in \cite{GGN1,GGN2}, count those
walks counted by the double Hurwitz numbers which have the additional property that the labels of the edges they 
traverse form a weakly increasing sequence of numbers. This is analogous to the relationship between simple
and self-interacting random walks.

	\begin{thm}[\cite{GGN3}]
	\label{thm:MainIZ}
		For any $1 \leq d \leq N$ and any $\alpha,\beta \in \Y^d$ we have
		
			\begin{equation*}
				I_N(\alpha,\beta) = (-1)^{\ell(\alpha)+\ell(\beta)}N^{-d}\sum_{g=-\infty}^\infty N^{2-2g} \mon_g^\bullet(\alpha,\beta)
			\end{equation*}
			
		\noindent
		and 
		
			\begin{equation*}
				L_N(\alpha,\beta) = (-1)^{\ell(\alpha)+\ell(\beta)}N^{-d} \sum_{g=0}^\infty N^{2-2g} \mon_g(\alpha,\beta),
			\end{equation*}
			
		\noindent
		where both series converge.
	\end{thm}
	
Theorem \ref{thm:MainIZ} gives a complete answer to the problem posed in \cite{IZ}, 
giving a complete genus expansion of the disconnected and connected string coefficients 
of the Ityzkson-Zuber integral. Concerning the analytic functions 

	\begin{equation}
		I_N = 1 + \sum_{d=1}^\infty \frac{z^d}{d!} I_N^d
		\quad\text{and}\quad
		L_N = \sum_{d=1}^\infty \frac{z^d}{d!} L_N^d
	\end{equation}
	
\noindent
themselves, Theorem \ref{thm:MainIZ} says that for all $1 \leq d \leq N$ we have

	\begin{equation}
	 	I_N^d = N^{-d}\sum_{g=-\infty}^\infty N^{2-2g} \sum_{\alpha,\beta \in \Y^d} 
		\frac{p_\alpha(a_1,\dots,a_N)}{(-N)^{\ell(\alpha)}} \frac{p_\beta(b_1,\dots,b_N)}{(-N)^{\ell(\beta)}}
		 \mon_g^\bullet(\alpha,\beta)
	\end{equation}
	
\noindent
and 

	\begin{equation}
	 	L_N^d = N^{-d}\sum_{g=0}^\infty N^{2-2g} \sum_{\alpha,\beta \in \Y^d} 
		\frac{p_\alpha(a_1,\dots,a_N)}{(-N)^{\ell(\alpha)}} \frac{p_\beta(b_1,\dots,b_N)}{(-N)^{\ell(\beta)}}
		 \mon_g(\alpha,\beta),
	\end{equation}
	
\noindent
where both series converge uniformly absolutely on compact subsets of $\C^{2N}$. Theorem \ref{thm:MainIZ}
thus indicates that as $N \to \infty$ we have

	\begin{equation}
	\label{eqn:FormalIZ}
		\log I_N \sim \sum_{g=0}^\infty N^{2-2g} F_g^\IZ,
	\end{equation}

\noindent
with 

	\begin{equation}
	\label{eqn:FreeEnergyIZ}
		F_g^\IZ = \sum_{d=1}^\infty \frac{z^d}{d!} N^{-d}\sum_{\alpha,\beta \in \Y^d} 
		\frac{p_\alpha(a_1,\dots,a_N)}{(-N)^{\ell(\alpha)}} \frac{p_\beta(b_1,\dots,b_N)}{(-N)^{\ell(\beta)}}
		 \mon_g(\alpha,\beta)
	\end{equation}
	
\noindent
a generating function for connected monotone Hurwitz numbers of genus $g$. 
The formal topological expansion \eqref{eqn:FormalIZ} of the Itzykson-Zuber free
energy is analogous to the formal topological expansion of the Hermitian one-matrix model
obtained in the classic papers \cite{BIPZ} and \cite{BIZ}, but with Hurwitz theory replacing
embedded graphs.

\subsection{Berezin-Karpelevich integral}
The purpose of this paper is to establish the counterpart of Theorem \ref{thm:MainIZ}
for the Berezin-Karpelevich integral

	\begin{equation}
	\label{eqn:BK}
		I_{MN} = \int_{\group{U}_M} \mathrm{d}U \int_{\group{U}_N}\mathrm{d}V e^{z \Tr (A^*UBV^* + VD^*U^*C)},
	\end{equation}
	
\noindent
where again $z \in \C$ is a coupling parameter but now $A,B,C,D \in \C^{M \times N}$ are 
complex rectangular matrices. By Fubini, we may assume $M \geq N$. The double integral $I_{MN}$ plays the role of $I_N$  
in the context of rectangular random matrices \cite{Benaych,GuiHua},
and we will see below that it depends on 
$A,B,C,D$ only up to the eigenvalues $x_1,\dots,x_N$ and $y_1,\dots,y_N$
of $A^*C,D^*B \in \C^{N \times N}$, so is again an entire function of $1+2N$
complex variables. Furthermore,
the Macluarin series of $I_{MN}$ can be presented as 

	\begin{equation}
	\label{eqn:StringExpansionBK}
		I_{MN} = 1 + \sum_{d=1}^\infty \frac{z^{2d}}{d!} \sum_{\alpha,\beta \in \Y^d} 
		\frac{p_\alpha(x_1,\dots,x_N)}{N^{\ell(\alpha)}} \frac{p_\beta(y_1,\dots,y_N)}{N^{\ell(\beta)}} I_{MN}(\alpha,\beta),
	\end{equation}
	
\noindent
which defines the string coefficients $I_{MN}(\alpha,\beta)$ of the Berezin-Karpelevich 
integral. The corresponding connected string expansion is

	\begin{equation}
		\label{eqn:ConnectedStringExpansionBK}
		\log I_{MN} = \sum_{d=1}^\infty \frac{z^{2d}}{d!} \sum_{\alpha,\beta \in \Y^d} 
		\frac{p_\alpha(x_1,\dots,x_N)}{N^{\ell(\alpha)}} \frac{p_\beta(y_1,\dots,y_N)}{N^{\ell(\beta)}}L_{MN}(\alpha,\beta),
	\end{equation}
	
\noindent
where $L_{MN}(\alpha,\beta)$ are the connected string coefficients of the Berezin-Karpelevich integral.

Our main result is an analogue of Theorem \ref{thm:MainIZ} giving a topological expansion of the string coefficients 
of the Berezin-Karpelevich integral and its logarithm in terms of a combinatorial refinement of monotone Hurwitz
numbers. Define the disconnected two-legged monotone Hurwitz number $\mon_g^\bullet(\alpha,\beta;s)$
to be the number of walks from a permutation of cycle type $\alpha$ to a permutation 
of cycle type $\beta$ in $r=2g-2+\ell(\alpha)+\ell(\beta)$ steps such that the labels 
of the edges traversed in the first $s$ steps are weakly increasing, as are the labels of
the edges traversed in the remaining $r-s$ steps. Thus,
we count walks with specified boundary conditions made up of two monotone
legs of specified length --- virtual histories of a self-interacting random 
walk on the symmetric group whose memory resets after a specified number of steps. 
As above, $\mon_g(\alpha,\beta;s)$ denotes the corresponding connected
Hurwitz number. 

	\begin{thm}
	\label{thm:MainBK}
		For any $1 \leq d \leq N$, and any $\alpha,\beta \in \Y^d$, we have
		
			\begin{equation*}
				I_{MN}(\alpha,\beta) = (-1)^{\ell(\alpha)+\ell(\beta)} (MN)^{-d} \sum_{g=-\infty}^\infty N^{2-2g} 
				\sum_{s=0}^{2g-2+\ell(\alpha)+\ell(\beta)} \left( \frac{N}{M} \right)^s \mon_g^\bullet(\alpha,\beta;s)
			\end{equation*}
			
		\noindent
		and 
		
			\begin{equation*}
				L_{MN}(\alpha,\beta) = (-1)^{\ell(\alpha)+\ell(\beta)} (MN)^{-d} \sum_{g=0}^\infty N^{2-2g} 
				\sum_{s=0}^{2g-2+\ell(\alpha)+\ell(\beta)} \left( \frac{N}{M} \right)^s \mon_g(\alpha,\beta;s),
			\end{equation*}
			
		\noindent
		where both series converge.
	\end{thm}
	
Just as Theorem \ref{thm:MainIZ} gives a formal topological expansion of the Ityzkson-Zuber integral, 
Theorem \ref{thm:MainBK} gives a formal large topological expansion of the Berezin-Karpelevich integral,

	\begin{equation}
	\label{eqn:FormalBK}
		\log I_{MN} \sim \sum_{g=0}^\infty N^{2-2g} F_g^{\BK}, \quad N \to \infty,
	\end{equation}
	
\noindent
in which the genus $g$ contribution 

	\begin{equation}
		F_g^\BK = \sum_{d=1}^\infty \frac{z^{2d}}{d!} (MN)^{-d} 
		\sum_{\alpha,\beta \in \Y^d} \frac{p_\alpha(x_1,\dots,x_N)}{(-N)^{\ell(\alpha)}} \frac{p_\beta(y_1,\dots,y_N)}{(-N)^{\ell(\beta)}}
		\sum_{s=0}^{2g-2+\ell(\alpha)+\ell(\beta)} \left( \frac{N}{M} \right)^s \mon_g(\alpha,\beta;s)
	\end{equation}
	
\noindent
is a generating function for two-legged monotone double Hurwitz numbers of genus $g$.
	
\section{Character Expansion}
In this section we derive the character expansion of the Berezin-Karpelevich integral, 
which is a known result \cite{GT1,GT2}. The conventional point of view is that
character expansions yield determinantal formulas, while in
our program they are antecedents of string expansions. We assume familiarity with the representation theory of the 
general linear and symmetric groups and globally cite \cite{Macdonald} as a
reference for this material.

	\subsection{Basic formulas}
	Isomorphism classes of irreducible polynomial representations of
	$\group{GL}_N = \Aut \C^N$ are parameterized by the set 
	$\Y_N$ of Young diagrams with at most $N$ rows. The character 
	
		\begin{equation}
			s_\lambda(A) = \Tr S^\lambda(A), \quad A \in \group{GL}_N,
		\end{equation}
		
	\noindent
	of any representative $(\W_N^\lambda,S^\lambda)$ of the class corresponding to $\lambda \in \Y_N$
	is a symmetric homogeneous polynomial function in the eigenvalues of $A$, 
	the Schur polynomial. We will write $s_\lambda(A)$ for the evaluation $s_\lambda(a_1,\dots,a_N)$ 
	of the Schur polynomial on the eigenvalues of any matrix $A \in \C^{N \times N}$.
	The pair $(\W_N^\lambda,S^\lambda)$ is an irreducible
	representation of $\group{U}_N \subset \group{GL}_N$, and we have the following 
	basic integration formulas.
	
		\begin{lem}
		\label{lem:Basic}
			For any Young diagrams $\lambda,\mu \in \Y_N$ and any matrices $X,Y \in \C^{N \times N}$,
			we have
			
				\begin{equation*}
					\int_{\group{U}_N} \mathrm{d}V s_\lambda(XVYV^*)  = \frac{s_\lambda(X)s_\lambda(Y)}{\dim \W_N^\lambda}
				\end{equation*}
				
			\noindent
			and 
			
				\begin{equation*}
					\int_{\group{U}_N} \mathrm{d}V s_\lambda(XV)s_\mu(YV^*) = 
					\delta_{\lambda\mu}\frac{s_\lambda(XY)}{\dim \W_N^\lambda}.
				\end{equation*}
		\end{lem}
		
		\begin{proof}
			Suppose first that $X,Y \in \group{GL}_N$. Then, $XUYU^* \in \group{GL}_N$ and we have
			
				\begin{equation*}
					s_\lambda(XUYU^*) = \Tr S^\lambda(X) S^\lambda(V) S^\lambda(Y) S^\lambda(V^*)
					= \sum_{i,j,k,l=1}^N S^\lambda(X)_{ij} S^\lambda(V)_{jk} S^\lambda(Y)_{kl} S^\lambda(V^*)_{li}.
				\end{equation*}
				
			\noindent
			We thus have 
			
				\begin{equation*}
				\int_{\group{U}_N} \mathrm{d}V s_\lambda(XUYU^*) 
				= \sum_{i,j,k,l=1}^N S^\lambda(X)_{ij} S^\lambda(Y)_{kl} \int_{\group{U}_N} \mathrm{d}VS^\lambda(V)_{jk} S^\lambda(V^*)_{li}.
				\end{equation*}
				
			\noindent
			By orthogonality of matrix elements in an irreducible representation of a compact group, we have
			
				\begin{equation*}
					\int_{\group{U}_N} \mathrm{d}VS^\lambda(V)_{jk} S^\lambda(V^*)_{li} = \frac{\delta_{ij}\delta_{kl}}{\dim \W_N^\lambda}, 
				\end{equation*}
				
			\noindent
			and thus
			
				\begin{equation*}
					\int_{\group{U}_N} \mathrm{d}Vs_\lambda(XVYV^*) = \frac{1}{\dim \W_N^\lambda}
					\left(\sum_{i=1}^N S^\lambda(X)_{ii} \right) \left(\sum_{k=1}^N S^\lambda(Y)_{kk} \right) = 
					\frac{1}{\dim \W_N^\lambda} \Tr S^\lambda(X) \Tr S^\lambda(Y).
				\end{equation*}
				
			We now extend to the case where $X,Y \in \C^{N \times N}$ are arbitrary matrices. 
			Write $f(V) = s_\lambda(XVYV^*)$. Since $\group{GL}_N$ is dense in $\C^{N \times N}$, 
			there are two sequences $(X_n)_{n=1}^\infty$ and $(Y_n)_{n=1}^\infty$ in 
			$\group{GL}_N$ such that 
			
				\begin{equation*}
					\lim_{n \to \infty} f_n(V) = f(V), \quad V \in \group{U}_N,
				\end{equation*}
				
			\noindent
			where $f_n(V) = s_\lambda(X_nVY_nV^*)$. Since the Schur polynomials are monomial positive, we have
			
				\begin{equation*}
					|f_n(V)| \leq s_\lambda(\|X_nVY_nV^*)\|,\dots,\|X_nVY_nV^*)\|) \leq \|X_n\|^d \|Y_n\|^d \dim \W_N^\lambda,
				\end{equation*}
				
			\noindent
			where $\|\cdot\|$ is operator norm and $d=|\lambda|$ is the number of cells in $\lambda$.
			Since $(\|X_n\|)_{n=1}^\infty$ and $(\|Y_n\|)_{n=1}^\infty$ are convergent sequences, they are
			bounded, and we may apply the dominated convergence theorem to conclude
			
				\begin{equation}
					\int_{\group{U}_N} \mathrm{d}V s_\lambda(XVYV^*) = \lim_{n \to \infty} \frac{s_\lambda(X_n) 
					s_\lambda(Y_n)}{\dim \W_n^\lambda} 
					= \frac{s_\lambda(X) s_\lambda(Y)}{\dim \W_n^\lambda}.
				\end{equation}
				
			The argument for the other integral is essentially the same, except that one uses orthogonality of 
			matrix elements in non-isomorphic irreducible representations.
				
		\end{proof}
		
		In addition to the above integration formulas, we will use Frobenius's formula for Schur
		polynomials in terms of Newton polynomials. Isomorphism classes of irreducible representations
		of $\group{S}^d=\Aut\{1,\dots,d\}$, or equivalently of the group algebra $\C\group{S}^d$, 
		are indexed by the set $\Y^d$ of Young diagrams with exactly
		$d$ cells. For each $\lambda \in \Y^d$, we choose a representative $(\V^\lambda,R^\lambda)$ of the
		irreducible representations corresponding to $\lambda \in \Y^d$. Moreover,
		for each $\alpha \in \Y^d$ we identity the conjugacy class $C_\alpha \subseteq \group{S}^d$
		of permutations of cycle type $\alpha$ with the formal sum of its elements in $\C\group{S}^d$.
		By Schur's Lemma, $R^\lambda(C_\alpha)$ is a scalar operator in $\End \V^\lambda$, and 
		we write $\omega_\alpha(\lambda)$ for its eigenvalue. The expansion of Schur polynomials
		on Newton polynomials is then
		
			\begin{equation}
			\label{eqn:Frobenius}
				s_\lambda = \frac{\dim \V^\lambda}{d!} \sum_{\alpha \in \Y^d} p_\alpha \omega_\alpha(\lambda),
				\quad \lambda \in \Y_N^d.
			\end{equation}

	\subsection{BGW character expansion}
	We now derive the character expansion of the Bars-Green/Br\'ezin-Gross-Witten/Wadia
	integral,
	
		\begin{equation}
		\label{eqn:BGW}
			J_N = \int_{\group{U}_N} \mathrm{d}V e^{z \Tr(X^*V + V^*Y)},
		\end{equation}
		
	\noindent
	the basic special function of $\group{U}_N$ gauge theory on a lattice of 
	any dimension \cite{BarsGreen,BG,GW,Samuel,Wadia}. This series expansion is due to Bars and Green \cite{BarsGreen}, 
	who considered the case where $X=Y$ and the action $V \mapsto \Tr(X^*V+V^*X)$ is real-valued.			
	
	\begin{thm}
	\label{thm:CharacterExpansionBGW}
		For any $z \in \C$ and $X,Y \in \C^{N \times N}$, we have
		
			\begin{equation*}
				J_N = 1 + \sum_{d=1}^\infty \frac{z^{2d}}{d!d!} \sum_{\lambda \in \Y_N^d} 
				s_\lambda(t_1,\dots,t_N) \frac{(\dim \V^\lambda)^2}{\dim \W_N^\lambda},
			\end{equation*} 
			
		\noindent
		where $t_1,\dots,t_N$ are the eigenvalues of $X^*Y$ and the series converges 
		absolutely. 
	\end{thm}
	
	\begin{proof}
	We view $J_N$ as an entire function of the complex variable $z$, with the 
	matrices $X,Y \in \C^{N \times N}$ arbitrary but fixed. The Maclaurin series of $J_N$ is
	
		\begin{equation}
			J_N = 1 + \sum_{d=1}^\infty \frac{z^{2d}}{d!d!}\int_{\group{U}_N} (\Tr X^*V)^d (\Tr YV^*)^d \mathrm{d}V,
		\end{equation}
		
	\noindent
	and 
	
		\begin{equation}
			(\Tr X^*V)^d (\Tr YV^*)^d = \left( \sum_{\lambda \in \Y_N^d} s_\lambda(X^*V) \dim \V^\lambda\right)
			\left( \sum_{\mu \in \Y_N^d} s_\mu(YV^*) \dim \V^\mu\right).
		\end{equation}
		
	\noindent
	The result now follows from Lemma \ref{lem:Basic}.
	\end{proof}

	\subsection{BK character expansion}	
	We now derive the character expansion of the Berezin-Karpelevich integral. 
	
		\begin{thm}
		\label{thm:CharacterExpansionBK}
			For any $z \in \C$ and $A,B,C,D \in \C^{M \times N}$, we have
			
				\begin{equation*}
					I_{MN} = 1 + \sum_{d=1}^\infty \frac{z^{2d}}{d!d!} 
					\sum_{\lambda \in \Y_N^d} s_\lambda(x_1,\dots,x_N) s_\lambda(y_1,\dots,y_N)
					\frac{\dim \V^\lambda}{\dim \W_M^\lambda}\frac{\dim \V^\lambda}{\dim \W_N^\lambda},
				\end{equation*}
				
			\noindent
			where $x_1,\dots,x_N$ are the eigenvalues of $A^*C$ and $y_1,\dots,y_N$ 
			are the eigenvalues of $D^*B$ and the series is absolutely convergent.
		\end{thm}
	
	\begin{proof}
	In the Berezin-Karpelevich integral \eqref{eqn:BK}, the inner integral over the 
	lower-rank unitary group,
		
		\begin{equation}
		\label{eqn:InnerIntegral}
			\int_{\group{U}_N} \mathrm{d}V e^{z\Tr (A^*UBV^* + VD^*U^*C)},
		\end{equation}
		
	\noindent
	is the BGW integral \eqref{eqn:BGW} with  
	
		\begin{equation}
		\label{eqn:BGWsubstitution}
			X = C^*UD \quad\text{ and }\quad Y =  A^*UB.
		\end{equation}
	
	\noindent
	Thus, by Theorem \ref{thm:CharacterExpansionBGW} the Maclaurin series of the Berezin-Karpelevich integral as a holomorphic
	function of $z$ is
	
		\begin{equation}
		\label{eqn:penultimate}
			I_{MN} =  \sum_{d=1}^\infty \frac{z^{2d}}{d!d!} \sum_{\lambda \in \Y_N^d}\frac{(\dim \V^\lambda)^2}{\dim \W_N^\lambda}
			\int_{\group{U}_M} \mathrm{d}U s_\lambda(D^*U^*CA^*UB).
		\end{equation}
		
	It remains to compute the integral
	
		\begin{equation}
			\int_{\group{U}_M} \mathrm{d}U s_\lambda(D^*U^*CA^*UB),
		\end{equation}
		
	\noindent
	where $D^*U^*CA^*UB \in \C^{N \times N}$. Recall that $M \geq N$.
	From the characteristic polynomial identity
	
		\begin{equation}
		\label{eqn:CharPolyIdentity}
			 x^{M-N}\det(xI_N-Z_1Z_2)=\det(xI_M-Z_2Z_1),
		\end{equation}
		
	\noindent
	which holds for arbitrary $Z_1 \in \C^{N \times M}$ and $Z_2 \in \C^{M \times N}$,
	the spectrum of $D^*U^*CA^*UB \in \C^{N \times N}$ coincides with that of $CA^*UBD^*U^* \in \C^{M \times M}$ 
	up to $M-N$ additional zero eigenvalues. Because the Schur polynomials are stable, 
	
		\begin{equation}
			s_\lambda(x_1,\dots,x_N)=s_\lambda(x_1,\dots,x_N,0,0,\dots,0),
		\end{equation}
	
	\noindent
	we have 
	
		\begin{equation}
			s_\lambda(C^*U^*DA^*UB) = s_\lambda(DA^*UBC^*U^*),
		\end{equation}
		
	\noindent
	and therefore
	
		\begin{equation}
			\int_{\group{U}_M} \mathrm{d}U s_\lambda(D^*U^*CA^*UB)  = \int_{\group{U}_M} \mathrm{d}U s_\lambda(CA^*UBD^*U^*).
		\end{equation}
	
	\noindent
	Note also that $\lambda \in \Y_N^d$ implies $\lambda \in \Y_M^d$ because
	$M \geq N$, so that $\lambda$ indexes an irreducible representation 
	$\W_M^\lambda$ of $\group{U}_M$, and
		
		\begin{equation}
			\int_{\group{U}_M} \mathrm{d}U s_\lambda(CA^*UBD^*U^*) = \frac{s_\lambda(CA^*)s_\lambda(BD^*)}{\dim \W_M^\lambda}
			= \frac{s_\lambda(A^*C)s_\lambda(D^*B)}{\dim \W_M^\lambda}.
		\end{equation}
	
	\end{proof}

	\section{Topological Expansion}
	In this Section we prove our main result, Theorem \ref{thm:MainBK}, which gives a topological
	expansion for the string coefficients and connected string coefficients of the Berezin-Karpelevich
	integral. We pair this with a topological expansion for the string coefficients of the 
	BGW integral.
	
		\subsection{String expansions}
		String expansions for the BGW integral $J_N$ and the Berezin-Karpelevich integral $I_{MN}$
		follow immediately from their character expansions, Theorems \ref{thm:CharacterExpansionBGW}
		and \ref{thm:CharacterExpansionBK}, together with Frobenius's formula \eqref{eqn:Frobenius}
		which expresses Schur polynomials in terms of Newton polynomials. 
		For the BGW integral, we have 
		
			\begin{equation}
			\label{eqn:StringExpansionBGW}
				J_N = 1 + \sum_{d=1}^\infty \frac{z^{2d}}{d!} \sum_{\alpha \in \Y^d} 
				\frac{p_\alpha(t_1,\dots,t_N)}{N^{\ell(\alpha)}} J_N(\alpha)
			\end{equation}
			
		\noindent
		with 
		
			\begin{equation}
			\label{eqn:StringCoefficientBGW}
				J_N(\alpha) = \frac{1}{d!}N^{\ell(\alpha)} \sum_{\lambda \in \Y_N^d} 
				\frac{(\dim \V^\lambda)^2}{d!} \omega_\alpha(\lambda) \frac{\dim \V^\lambda}{\dim \W_N^\lambda}.
			\end{equation}
			
		\noindent
		For the Berezin-Karpelevich integral, we obtain
		
			\begin{equation}
			\label{eqn:StringExpansionBK2}
			I_{MN} = 1 + \sum_{d=1}^\infty \frac{z^{2d}}{d!} \sum_{\alpha,\beta \in \Y^d} 
			\frac{p_\alpha(x_1,\dots,x_N)}{N^{\ell(\alpha)}} \frac{p_\beta(y_1,\dots,y_N)}{N^{\ell(\beta)}} I_{MN}(\alpha,\beta)
			\end{equation}
			
		\noindent
		with 
		
			\begin{equation}
			\label{eqn:StringCoefficientBK}
				I_{MN}(\alpha,\beta) = \frac{1}{d!d!}N^{\ell(\alpha)+\ell(\beta)} \sum_{\lambda \in \Y_N^d} 
				\frac{(\dim \V^\lambda)^2}{d!} 
				\omega_\alpha(\lambda) \frac{\dim \V^\lambda}{\dim \W_M^\lambda}\frac{\dim \V^\lambda}{\dim \W_N^\lambda}\omega_\beta(\lambda).
			\end{equation}
		
		Using the standard dimension formulas \cite{Macdonald}, for any $\lambda \in \Y_N^d$ we have
		
			\begin{equation}
			\label{eqn:DimensionRatio}
				\frac{\dim \V^\lambda}{\dim \W_N^\lambda} = \frac{d!}{N^d} \Omega_{\frac{1}{N}}^{-1}(\lambda),
			\end{equation}

		\noindent
		with
		
			\begin{equation}
				\Omega_\hbar(\lambda) = \prod_{\Box \in \lambda}(1+\hbar c(\Box))
			\end{equation}
			
		\noindent
		the \emph{content polynomial} of $\lambda$, in which $c(\Box)$ is the column index minus the row index of a given cell $\Box \in \lambda$.
		Note that $\Omega_\frac{1}{N}(\lambda)>0$ for any $\lambda \in \Y_N$.
		Using \eqref{eqn:DimensionRatio} the BGW and Berezin-Karpelevich string coefficients become
		
			\begin{equation}
			\label{eqn:StringCoefficientContentBGW}
				J_N(\alpha) = N^{\ell(\alpha)-d} \sum_{\lambda \in \Y_N^d} 
				\frac{(\dim \V^\lambda)^2}{d!} \omega_\alpha(\lambda) \Omega_\frac{1}{N}^{-1}(\lambda)
			\end{equation}
			
		\noindent
		and 
		
			\begin{equation}
			\label{eqn:StringCoefficientContentBK}
				I_{MN}(\alpha,\beta) = M^{-d}N^{\ell(\alpha)+\ell(\beta)-d} \sum_{\lambda \in \Y_N^d} 
				\frac{(\dim \V^\lambda)^2}{d!} 
				\omega_\alpha(\lambda) \Omega_\frac{1}{M}^{-1}(\lambda)\Omega_\frac{1}{N}^{-1}(\lambda)\omega_\beta(\lambda).
			\end{equation}
		
		\noindent
		These formulas are already enough to obtain $1/N$ expansions for the string coefficients of the 
		BGW and Berezin-Karpelevich integrals.
				
			\begin{prop}
			\label{prop:ContentExpansionBGW}
				For any Young diagram $\alpha$ with $d \leq N$ cells, we have
								
					\begin{equation*}
						J_N(\alpha) = N^{\ell(\alpha)-d} \sum_{r=0}^\infty \frac{(-1)^r}{N^r} \sum_{\lambda \in \Y_N^d} \frac{(\dim \V^\lambda)^2}{d!} 
						\omega_\alpha(\lambda) h_r(\lambda),
					\end{equation*}
					
				\noindent
				where the series converges absolutely and $h_r(\lambda)$ denotes the evaluation 
				of the complete symmetric polynomial of degree $r$ on the multiset of contents of $\lambda$.
			\end{prop}
			
			\begin{proof}
				The contents of any Young diagram $\lambda$ contained in the $N \times N$ square diagram
				are all strictly less than $N$ in absolute value. Thus, for any such diagram we have
				
					\begin{equation*}
						\Omega_\frac{1}{N}^{-1}(\lambda) = \prod_{\Box \in \lambda} 
						\left(1+\frac{c(\Box)}{N}\right)^{-1}= \sum_{r=0}^\infty \frac{(-1)^r}{N^r} h_r(\lambda),
					\end{equation*}
					
				\noindent
				where the series is absolutely convergent, and plugging this absolutely convergent
				expansion into \eqref{eqn:StringCoefficientContentBGW} yields the result.
			\end{proof}
			
			\begin{prop}
			\label{prop:ContentExpansionBK}
				For any Young diagrams $\alpha,\beta$ with $d \leq N$ cells, 
				we have
				
					\begin{equation*}
						I_{MN}(\alpha,\beta) = M^{-d} N^{\ell(\alpha)+\ell(\beta)-d} \sum_{r=0}^\infty \frac{(-1)^r}{N^r} \sum_{s=0}^r v^s 
						\sum_{\lambda \in \Y_N^d} \frac{(\dim \V^\lambda)^2}{d!} 
						\omega_\alpha(\lambda) h_s(\lambda) h_{r-s}(\lambda) \omega_\beta(\lambda),
					\end{equation*} 
					
				\noindent
				where the series converges absolutely.
			\end{prop}
			
			\begin{proof}
				The proof is the same as the proof of the preceding proposition, except that we have the double 
				product 
				
					\begin{equation*}
						\Omega_\frac{1}{M}^{-1}(\lambda) \Omega_\frac{1}{N}^{-1}(\lambda) 
						= \prod_{\Box \in \lambda} \left(1+\frac{c(\Box)}{M}\right)^{-1}
						\left(1+\frac{c(\Box)}{N}\right)^{-1},
					\end{equation*}
					
				\noindent
				which we write as 
				
					\begin{equation*}
						\Omega_\frac{1}{M}^{-1}(\lambda) \Omega_\frac{1}{N}^{-1}(\lambda) 
						= \prod_{\Box \in \lambda} \left(1+\frac{vc(\Box)}{N}\right)^{-1}
						\left(1+\frac{c(\Box)}{N}\right)^{-1}
					\end{equation*}
					
				\noindent
				with $v = \frac{N}{M} \leq 1$. Then, for any Young diagram $\lambda$ contained in 
				the $N \times N$ square we have
				
					\begin{equation*}
						\Omega_\frac{1}{M}^{-1}(\lambda) \Omega_\frac{1}{N}^{-1}(\lambda)
						= \sum_{r_1,r_2=0}^\infty \frac{(-v)^{r_1}}{N^{r_1}} \frac{(-1)^{r_2}}{N^{r_2}}
						h_{r_1}(\lambda)h_{r_2}(\lambda) = \sum_{r=0}^\infty \frac{(-1)^r}{N^r} 
						\sum_{s=0}^r v^r h_s(\lambda) h_{r-s}(\lambda),
					\end{equation*}
					
				\noindent
				where the series converges absolutely and can be substituted into 
				\eqref{eqn:StringCoefficientContentBK}.
			\end{proof}
						
	\subsection{Disconnected topological expansion}
	The Fourier transform gives an algebra isomorphism from the center of $\C\group{S}^d$
	to the pointwise algebra of complex-valued functions on Young diagrams:
	if $A \in \C\group{S}^d$ is a central function, its Fourier transform $\hat{A}(\lambda)$ is the unique
	eigenvalue of the scalar operator $R^\lambda(A)$ acting in $\V^\lambda$. Furthermore,
	the canonical trace $\langle \cdot \rangle$ on $\C\group{S}^d$, i.e. the normalized character
	of the regular representation, is for central elements implemented by the Plancherel formula
	
		\begin{equation}
			\langle A \rangle = \sum_{\lambda \in \Y^d} \frac{(\dim \V^\lambda)^2}{d!} \hat{A}(\lambda).
		\end{equation}
		
	By definition, $\omega_\alpha(\lambda)$ is the Fourier transform of a conjugacy class,
		
		\begin{equation}
			\hat{C}_\alpha(\lambda) = \omega_\alpha(\lambda).
		\end{equation}
		
	\noindent
	The theorem of Jucys \cite{Jucys} and Murphy \cite{Murphy}
	says that any symmetric polynomial $f(J_1,\dots,J_d)$ in the Jucys-Murphy elements 
	
		\begin{equation}
			J_j = \sum_{i=1}^j (i\ j), \quad 1 \leq j \leq d,
		\end{equation}
		
	\noindent
	is a central element in $\C\group{S}^d$, and that its Fourier transform 
	is the function $f(\lambda)$ obtained by evaluation of $f$ on the multiset of 
	contents of $\lambda$. We thus have that 
	
		\begin{equation}
			\langle C_\alpha h_r \rangle =  \sum_{\lambda \in \Y^d} \frac{(\dim \V^\lambda)^2}{d!} 
			\omega_\alpha(\lambda) h_r(\lambda)
		\end{equation}
		
	\noindent
	is the coefficient of the identity permutation in the product $C_\alpha h_r(J_1,\dots,J_d)$,
	which is precisely the number $\vec{W}^r(\alpha)$ of monotone $r$-step walks from the 
	identity permutation to a permutation of cycle type $\alpha$ on the Cayley graph of
	$\group{S}^d$. Thus, Proposition \ref{prop:ContentExpansionBGW} implies that for 
	any Young diagram $\alpha$ with $d \leq N$ cells, we have
	
		\begin{equation}
				J_N(\alpha) = N^{\ell(\alpha)-d} \sum_{r=0}^\infty \frac{(-1)^r}{N^r} \vec{W}^r(\alpha),		
		\end{equation}
		
	\noindent
	the series being absolutely convergent. That is, the string coefficients of the BGW integral
	are generating functions enumerating monotone walks of specified length from the 
	identity to a specified conjugacy class. Similarly,
	
		\begin{equation}
			\langle C_\alpha h_s h_{r-s} C_\beta \rangle =  \sum_{\lambda \in \Y^d} \frac{(\dim \V^\lambda)^2}{d!} 
			\omega_\alpha(\lambda) h_s(\lambda)h_{r-s}(\lambda) \omega_\beta(\lambda)
		\end{equation}
		
	\noindent
	is the number $\vec{W}^r(\alpha,\beta;s)$ of $r$-step walks $C_\alpha \to C_\beta$ on 
	$\group{S}^d$ consisting of two monotone legs, one of length $s$ followed by 
	one of length $r-s$. Therefore Proposition \ref{prop:ContentExpansionBK} is 
	equivalent to the statement that for all $\alpha,\beta$ with $d \leq N$ cells we 
	have
	
		\begin{equation}
		\label{eqn:PreRH}
			I_{MN}(\alpha,\beta) = M^{-d} N^{\ell(\alpha)+\ell(\beta)-d} \sum_{r=0}^\infty \frac{(-1)^r}{N^r} \sum_{s=0}^r v^s 
					\vec{W}^r(\alpha,\beta;s),
		\end{equation} 
		
	\noindent
	so that the string coefficients of the Berezin-Karpelevich integral 
	are generating functions for two-legged monotone walks of specified 
	length between specified conjugacy classes.
	
	According to the Riemann-Hurwitz formula, $\vec{W}^r(\alpha)$ vanishes unless 
	$r = 2g-2+\ell(\alpha) +d$, where $|\alpha|=d$. Thus, we obtain the following 
	genus expansion for the string coefficients of the BGW integral.

		\begin{thm}
		\label{thm:DisconnectedTopologicalBGW}
			For any Young diagram $\alpha$ with $d \leq N$ cells, we have
			
				\begin{equation*}
					J_N(\alpha) =(-1)^{\ell(\alpha)+d} N^{-2d} \sum_{g=-\infty}^\infty N^{2-2g} \mon_g^\bullet(\alpha),
				\end{equation*}
				
			\noindent
			where $\mon_g^\bullet(\alpha)= \vec{W}^{2g-2+\ell(\alpha)+d}(\alpha)$ is the 
			disconnected monotone single Hurwitz number of genus $g$, and the series converges.
		\end{thm}
		
	Applying the Riemann-Hurwitz formula in \eqref{eqn:PreRH}, we likewise 
	obtain a genus expansion for the string coefficients of the Berezin-Karpelevich integral.
		
		\begin{thm}
		\label{thm:DisconnectedTopologicalBK}
			For any Young diagrams $\alpha,\beta$ with $d \leq N$ cells, we have
			
				\begin{equation*}
				I_{MN}(\alpha,\beta) = (-1)^{\ell(\alpha)+\ell(\beta)} (MN)^{-d} \sum_{g=-\infty}^\infty N^{2-2g} 
				\sum_{s=0}^{2g-2+\ell(\alpha)+\ell(\beta)} v^s \mon_g^\bullet(\alpha,\beta;s)
			\end{equation*}
			
			\noindent
			where $\mon_g^\bullet(\alpha,\beta;s) = \vec{W}^{2g-2+\ell(\alpha)+\ell(\beta)}(\alpha,\beta;s)$ is the disconnected two-legged
			monotone double Hurwitz number of genus $g$, and the series converges.
		\end{thm}
		
	\subsection{Connected topological expansion}
	The BGW integral $J_N$ is an analytic function on 
	$\C^{1+N}$, and its logarithm $\log J_N$ is analytic 
	on an open neighborhood of the origin. Writing 
	
		\begin{equation}
			J_N = 1 + \sum_{d=1}^\infty \frac{z^{2d}}{d!} J_N^d \quad\text{and}\quad
			\log J_N = 1 + \sum_{d=1}^\infty \frac{z^{2d}}{d!} K_N^d
		\end{equation}

	\noindent		
	each coefficient $K_N^d$ is a polynomial in $J_N^1,\dots,J_N^d$ given explicitly by 
	the Exponential Formula \cite{GJ:Book}.
	This in turn gives an explicit relation between the string coefficients $J_N(\alpha)$ of $J_N$ and the 
	string coefficients $K_N(\alpha)$ of its logarithm, which are defined by the expansion
	
		\begin{equation}
			K_N^d = \sum_{\alpha \in \Y^d}
			\frac{p_\alpha(t_1,\dots,t_N)}{N^{\ell(\alpha)}} K_N(\alpha).
		\end{equation}
	
	\noindent
	As a consequence this \cite{GGN4}, Theorem \ref{thm:DisconnectedTopologicalBGW}
	is equivalent to the following result.
	
		\begin{thm}
		\label{thm:ConnectedTopologicalBGW}
			For any Young diagram $\alpha$ with $d \leq N$ cells, we have
			
				\begin{equation*}
					J_N(\alpha) =(-1)^{\ell(\alpha)+d} N^{-2d} \sum_{g=0}^\infty N^{2-2g} \mon_g(\alpha),
				\end{equation*}
				
			\noindent
			where $\mon_g(\alpha)$ is the 
			monotone single Hurwitz number of genus $g$, and the series converges.
		\end{thm}

	An explicit formula for $\mon_0(\alpha)$ is given in \cite{GGN1}, and it is equivalent to 
	the first-order asymptotics of the BGW integral derived by O'Brien and Zuber \cite{OZ1,OZ2}
	in the context of lattice gauge theory; see also \cite{GrossNew}. Going further, an explicit formula for $\mon_1(\alpha)$
	is given in \cite{GGN2}, so that the genus one correction to the formula of O'Brien and Zuber
	follows from Theorem \ref{thm:ConnectedTopologicalBGW}. Likewise, applying the moment-cumulant
	formula to Theorem \ref{thm:DisconnectedTopologicalBK} completes the proof of 
	Theorem \ref{thm:MainBK}, giving a topological expansion for both the string coefficients
	and connected string coefficients of the Berezin-Karpelevich integral.
		
	\section{Combinatorial Identities from Matrix Integrals}
	
		\subsection{Itzykson-Zuber case}
		Taking $B$ to be the identity matrix in the Itzykson-Zuber integral, we obtain 
		an exponential function: in this specialization 
		
			\begin{equation}
			\label{eqn:DegenerationIZ}
				\log I_N = z(a_1+\dots+a_N).
			\end{equation}
			
		\noindent
		Together with Theorem \ref{thm:MainIZ}, this degeneration of $I_N$ implies
		the following cancellation identity for monotone double Hurwitz numbers.
		
			\begin{thm}
			\label{thm:Cancellation}
				For any $(d,g) \in \N \times \N_0$ except $(1,0)$, 
				we have
				
					\begin{equation*}
						\sum_{\beta \in \Y^d} (-1)^{\ell(\alpha)} \mon_g(\alpha,\beta) = 0
					\end{equation*}
					
				\noindent
				for all $\alpha \in \Y^d$.
			\end{thm}
			
			\begin{proof}
				The case where $d=1$ and $g>0$ is combinatorially obvious: the sum consists of 
				the single term $\mon_g(1,1)$, which vanishes as there are no walks of positive
				length in a graph with a single vertex.
			
				For $d > 1$, the result is a consequence of Theorem \ref{thm:MainIZ} together with the 
				degeneration \eqref{eqn:DegenerationIZ}. More precisely, writing 
				
					\begin{equation}
						\log I_N = \sum_{d=1}^\infty \frac{z^d}{d!} L_N^d,
					\end{equation}
					
				\noindent
				in the degeneration \eqref{eqn:DegenerationIZ} we have that $L_N^d=0$ for all $d >1$.
				Let $d >1$ be arbitrary but fixed. Then by Theorem \ref{thm:MainIZ}, for all $N \geq d$ we have that
				
					\begin{equation}
						\sum_{\alpha,\beta \vdash d} \frac{p_\alpha(a_1,\dots,a_N)}{(-N)^{\ell(\alpha)}} (-1)^{\ell(\beta)}
						\sum_{g=0}^\infty N^{2-2g} \mon_g(\alpha,\beta) = 0,
					\end{equation}
					
				\noindent
				which forces
				
					\begin{equation}
					\label{eqn:InfiniteCancellation}
						\sum_{\beta \in \Y^d} (-1)^{\ell(\beta)} \sum_{g=0}^\infty N^{-2g} \mon_g(\alpha,\beta) =
						 \sum_{g=0}^\infty N^{-2g} \sum_{\alpha \beta \in \Y^d} (-1)^{\ell(\beta)} \mon_g(\alpha,\beta)= 0
					\end{equation}
					
				\noindent
				for each $\alpha \in \Y^d$ and all $N \geq d$, by linear independence of the degree $d$ Newton polynomials 
				in $N \geq d$ variables.
								
				We now proceed by induction in $g$. For $g=0$, take the $N \to \infty$ limit 
				in \eqref{eqn:InfiniteCancellation} to obtain
				
					\begin{equation}
						\sum_{\beta \in \Y^d} (-1)^{\ell(\beta)} \mon_0(\alpha,\beta)=0
					\end{equation}
					
				\noindent
				for each $\alpha \in \Y^d$. Assuming the result holds up to genus $k$, 
				\eqref{eqn:InfiniteCancellation} becomes 
				
					\begin{equation}
					\label{eqn:InfiniteCancellationInduction}
						\sum_{g=k+1}^\infty N^{-2g} \sum_{\beta \in \Y^d} (-1)^{\ell(\beta)} \mon_g(\alpha,\beta) = 0,
					\end{equation}
					
				\noindent
				for each $\alpha \in \Y^d$.
				Multiply \eqref{eqn:InfiniteCancellationInduction} by $N^{2k}$ and take the $N \to \infty$ limit to 
				obtain
				
					\begin{equation}
						\sum_{\beta \in \Y^d} (-1)^{\ell(\beta)} \mon_{k+1}(\alpha,\beta)=0
					\end{equation}
					
				\noindent
				for each $\alpha \in \Y^d$.
					
			\end{proof}

		\subsection{Berezin-Karpelevich case}
		We now consider an analogous specialization of the Berezin-Karpelevich integral
		which, via Theorem \ref{thm:MainBK}, produces a combinatorial identity for two-legged
		monotone Hurwitz numbers. We consider the case of equal dimensions, $M=N$, and 
		take the matrices $B$ and $D$ to be the identity. In this specialization we see that, by invariance of
		Haar measure, the Berezin-Karpelevich integral degenerates to the BGW integral,

			\begin{equation}
				I_{NN} = \int_{\group{U}_N} \mathrm{d}U \int_{\group{U}_N} \mathrm{d}V 
				e^{z\Tr (A^*UV^*+ VU^*C)} = \int_{\group{U}_N} \mathrm{d}U e^{z(\Tr A^*U + U^*C)} = J_N.
			\end{equation}
			
		\noindent
		Thus, for each $d \in \N$ we have the polynomial identity   
		
			\begin{equation}
				\sum_{\alpha,\beta \in \Y^d} \frac{p_\alpha(x_1,\dots,x_N)}{N^{\ell(\alpha)}} L_{NN}(\alpha,\beta) =
				\sum_{\alpha \in \Y^d} \frac{p_\alpha(x_1,\dots,x_N)}{N^{\ell(\alpha)}} K_N(\alpha),
			\end{equation}
			
		\noindent
		where $x_1,\dots,x_N$ are the eigenvalues of $A^*C \in \C^{N \times}$,
		and $L_{NN}(\alpha,\beta)$ and $K_N(\alpha)$ are the connected string coefficients
		of $I_{NN}$ and $J_N$, respectively, which mplies that for any 
		$\alpha \in \Y^d$ and all $N \geq d$ we have the numerical identity 
		
			\begin{equation}
				\sum_{\beta \in \Y^d} L_{NN}(\alpha,\beta) = K_N(\alpha).
			\end{equation}
			
		\noindent
		By Theorems \ref{thm:MainBK} and \ref{thm:DisconnectedTopologicalBGW}, this in turn gives
		
			\begin{equation}
				 \sum_{g=0}^\infty N^{2-2g} 
				\sum_{\beta \in \Y^d} (-1)^{\ell(\beta)}\sum_{s=0}^{2g-2+\ell(\alpha)+\ell(\beta)} 
				\mon_g(\alpha,\beta;s) = (-1)^d \sum_{g=0}^\infty N^{2-2g} \mon_g(\alpha),
			\end{equation}
			
		\noindent
		for each $\alpha \in \Y^d$ and all $N \geq d$, both series being convergent. We thus obtain
		the following summation formula, which is useful in the theory of the rectangular $R$-transform
		\cite{McNovakFree}.
	
		\begin{thm}
			For any degree $d \in \N$, genus $g \in \N_0$, and profile 
			$\alpha \in \Y^d$, we have
			
				\begin{equation*}
					\sum_{\beta \in \Y^d} (-1)^{\ell(\beta)} \sum_{s=0}^{2g-2+\ell(\alpha)+\ell(\beta)} \mon_g(\alpha,\beta;s) =
					(-1)^d \mon_g(\alpha).
				\end{equation*}
		\end{thm}
		
		\subsubsection{Conflict of interest statement}
		The author states that there is no conflict of interest.
		
		\subsubsection{Data avaialbility statement}
		The author states that all relevant data is included.

\end{document}